\documentclass[a4paper,fleqn,usenatbib]{mnras}

\usepackage{times}

\usepackage[T1]{fontenc}
\usepackage{ae,aecompl}

\usepackage{graphicx}	
\usepackage{amsmath}	
\usepackage{amssymb}	
\usepackage{color}
\usepackage{multirow}
\usepackage{hyperref}
\title[Radio re-brightenings in Sw J0243]{A re-establishing jet during an X-ray re-brightening of \\ the Be/X-ray binary Swift J0243.6+6124}

\author[Van den Eijnden et al.]{
\noindent J. van den Eijnden$^{1}$\thanks{E-mail: a.j.vandeneijnden@uva.nl},
N. Degenaar$^{1}$,
T. D. Russell$^{1}$,
J. V. Hern\'andez Santisteban$^{1}$,
\newauthor R. Wijnands$^{1}$, 
J. C. A. Miller-Jones$^{2}$,
A. Rouco Escorial$^{1}$,
G. R. Sivakoff$^{3}$\\
$^{1}$Anton Pannekoek Institute for Astronomy, University of Amsterdam, Science Park 904, 1098 XH Amsterdam, The Netherlands\\
$^{2}$International Centre for Radio Astronomy Research -- Curtin University, GPO Box U1987, Perth, WA 6845, Australia\\
$^{3}$Department of Physics, University of Alberta, CCIS 4-183, Edmonton, AB T6G 2E1, Canada\\}

\date{Accepted XXX. Received YYY; in original form ZZZ}

\pubyear{2018}

\begin{document}
\label{firstpage}
\pagerange{\pageref{firstpage}--\pageref{lastpage}}
\maketitle

\begin{abstract}
\noindent Transient Be/X-ray binary systems, wherein a compact object accretes from a Be-companion star, can show giant and periastron outbursts. During the decay of their giant outbursts, some Be/X-ray binaries also show X-ray re-brightenings, the origin of which is not understood. Recently, we presented the discovery of a jet from a neutron star Be/X-ray binary, observed during the giant outburst of Swift J0243.6+6124. Here, we present continued radio monitoring of its 2017/2018 giant outburst decay and a re-brightening of this source. During the former, we observe a radio flare with a steep radio spectrum, possibly caused by interactions between discrete ejecta colliding with the pre-existing jet or the surrounding medium. During the X-ray re-brightening, we observe the radio jet turning on and off within days. Surprisingly, this re-establishing jet is as bright in radio as at the peak of the super-Eddington giant outburst, despite more than two orders of magnitude lower X-ray luminosity. In addition, the jet is only observed when the X-ray luminosity exceeds approximately $2\times 10^{36} (D/5\rm kpc)^2$ erg/s. We discuss how such an X-ray threshold for jet launching might be related to the presence of a magnetic centrifugal barrier at lower mass accretion rates. We also discuss the implications of our results for the launch of jets from strongly magnetized neutron stars, and explore future avenues to exploit the new possibility of coordinated X-ray/radio studies of neutron star Be/X-ray binaries. 
\end{abstract}




\section{Introduction}

X-ray binaries, wherein either a black hole (BH) or a neutron star (NS) accretes gas from a stellar binary companion, are often classified based on the mass of the donor star. In low-mass X-ray binaries (LMXBs), the companion's mass is typically below $1 M_{\odot}$ and mass transfer predominantly occurs through Roche-lobe overflow \citep{shakura73}. In high-mass X-ray binaries (HMXBs), the mass of the donor is (much) larger, often exceeding that of the accretor, and mass transfer occurs primarily through stellar winds or a decretion disk \citep{reig11}. A significant fraction of Galactic X-ray binaries are HMXBs, which generally host a NS accretor \citep{liu06,liu07}; these NSs in HMXBs in turn have strong ($B \gtrsim 10^{12}$ G) magnetic fields and long spin periods \citep{shi15}. On the contrary, NSs in LMXBs typically have much weaker inferred magnetic fields, below $\sim 10^{8-9}$ G \citep{mukherjee15}, and faster (e.g. $>> 1$ Hz) spin frequencies \citep{patruno12}. However, to complicate this picture, a small number of LMXBs contain strongly magnetized, slowly spinning NSs instead.  \citep[e.g.][]{rappaport77,kouveliotou96,jonker01,bahramian14,kuranov15}.  

The HMXBs hosting a NS can be further divided into two main subclasses, based on their donor \citep[see][for a detailed review]{reig11}: the supergiant X-ray binaries and the Be/X-ray binaries (BeXRBs). In the latter, on which we focus in this work, the donor is a Be-star \citep{slettebak88, balona00, porter03}. Such sources can show two main types of transient outbursts \citep{stella86}: short-lived periastron or \textit{Type-I} outbursts, where the NS moves through the circumstellar decretion disk of the Be donor at periastron passage \citep{okazaki01}, and more prolonged giant or \textit{Type-II} outbursts, unrelated to any specific orbital phase \citep{moritani13, martin14, monageng17, laplace17}. The latter can last multiple orbital periods and reach (super-)Eddington accretion rates, while the former tend to reach X-ray luminosities of only $\sim 10^{36}$ -- $10^{37}$ erg/s. During the decay of a giant outburst, re-brightenings are sometimes observed in X-rays \citep{chen97,wijnands16,tsygankov16a,sguera17,rouco17}. These events, sometimes referred to as re-flares, might be related to periastron passages -- similar to Type-I outbursts -- or, since not all re-brightenings occur at periastron, to cooling waves moving through the accretion flow \citep{tsygankov17}. Their exact origin remains unclear and may differ between objects and outbursts. 

During an outburst, transient X-ray binaries can launch part of the material in highly collimated outflows called jets \citep[see e.g.][for overviews on both black holes and neutron stars]{fender04,migliari06,fender14,fender16}. Emitting primarily through synchrotron processes, these jets can dominate X-ray binary spectra from radio up to (mid-)infrared frequencies \citep[][]{russellD06,russellD07,russell14,diaztrigo18}. Jets can also contribute up to X-ray energies through inverse Compton processes \citep[see e.g.][]{markoff03}. These jets offer an independent view of the processes in the X-ray binary, complementary to the X-rays emitted by the infalling material. For instance, studying the correlation between the radio and X-ray lumunosity has revealed a clear coupling between the accretion and ejection processes, in both black hole and neutron star systems \citep[see e.g.][]{corbel03,gallo03,migliari03,gusinskaia17,tudor17,gallo18}. 

However, until recently, jets were never observed in either LMXBs or HMXBs with a strongly-magnetized NS accretor \citep{wright75,nelson88,fender00,migliari06,migliari12}. While the formation of jets in such systems is not well understood, it was long thought that their strong magnetic field inhibited the launch of such outflows \citep[e.g.][]{blandford82,massi08}. As HMXB NSs typically have strong magnetic fields, this paradigm implied that those binaries have rarely been studied from the perspective of comparing the radio and X-ray emission, connecting the in- and outflow of material. 

The first observational hints that NSs with magnetic fields exceeding $\sim 10^{12}$ G might be able to launch jets were the detections of radio emission of unconfirmed origin from the accreting X-ray pulsars GX 1+4 and Her X-1 \citep{vandeneijnden18_her,vandeneijnden18_gx}. More recently, in \citet{vandeneijnden18_nature} we reported the first jet detection from a strongly-magnetized accreting neutron star. This jet was observed with the Karl G. Jansky Very Large Array (VLA) during a giant, super-Eddington outburst of the BeXRB Swift J0243.6+6124 (hereafter Sw J0243) -- the first outburst of this transient, which was discovered in October 2017 \citep{kennea17}. This detection clearly shows that NS BeXRBs are capable of launching jets, opening a new avenue to study the accretion process in these systems. Additionally, it opens up a new parameter space of jet formation, with strong magnetic fields and slow accretor spins. 

Here, we report on continued VLA monitoring of Sw J0243 during the decay of the giant outburst. In this paper, we will discuss both the implications for the physics of jets from strongly-magnetized NSs, and showcase the potential of understanding jet formation and X-ray re-brightenings in BeXRBs through coordinated X-ray and radio monitoring. 

\section{Observations}

\subsection{Radio}

After the discovery of the first outburst of Sw J0243 and the following detection of $9.8$ sec X-ray pulsations \citep{kennea17}, identifying the accretor as a slowly spinning NS, we monitored the source with the VLA over $15$ epochs (spanning $\sim$ eight months). The first $8$ epochs, focussing on the peak of the giant outburst, showed the onset and evolution of a jet and are fully discussed in \citet{vandeneijnden18_nature}. Here, we focus on the remaining $7$ epochs (VLA project codes 17B-420 and 18A-456, with the VLA in configurations A and C, respectively), which targeted the late decay and the re-brightenings of the outburst. All observing dates/MJDs are listed in Table \ref{tab:obs}. 

In Table \ref{tab:obs}, we also list the observing frequencies per epoch. We observed Sw J0243 at different (combinations) of frequencies in C and K band, to optimize both frequency coverage and sensitivity per observing band. In C band, we first observed in 3-bit mode, providing the maximum bandwidth ($4-8$ GHz, epochs 1 and 2), and subsequently switched to 8-bit mode, with two simultaneous bands centred at $4.5$ and $7.5$ GHz with $1$ GHz of bandwidth each (epochs 3 to 7). In epoch 2, we included K band in 3-bit mode, centred at $22$ GHz with $8$ GHz of bandwidth. During all observations, J0137+3309 (3C 48) was the primary calibrator, and the nearby ($\sim 1^{\rm o}$ away) phase calibrator was J0244+6228. 

We followed the same data analysis method as described in \citet{vandeneijnden18_nature} for the first eight epochs: we analysed the observations using the \textsc{Common Astronomy Software Application} \citep[\textsc{casa};][]{mcmullin07} package v4.7.2, removing RFI using both automatic flagging routines and careful visual inspection. After calibration, we imaged Stokes I using the multi-scale multi-frequency \textsc{clean} task, setting the robust weighting parameter to one. We then measured flux densities by fitting an elliptical Gaussian equal to the beam size with the \textsc{casa}-task \textsc{imfit}. The RMS was determined over a nearby area without any sources. When Sw J0243 was not detected (epochs 3, 4, and 7), we combined the $4.5$ and $7.5$ GHz bands to obtain the highest sensitivity. We then measured the RMS over the source position and tripled this value to determine the $3\sigma$ upper limit. 

\subsection{X-rays}
\label{sec:xray_methods}

To track the accretion state and outburst evolution of Sw J0243, we determined X-ray fluxes during each radio epoch. During the first three VLA epochs, the Neil Gehrels Swift Observatory (hereafter Swift) observed Sw J0243 on the same day with the on-board X-ray Telescope (XRT), providing a quasi-simultaneous X-ray spectrum. However, between April 29 and June 19, 2018, Swift did not observe Sw J0243 due to Sun constraints. Therefore, we instead used the Monitor of All-sky X-ray Image (MAXI) Gas Slit Camera (GSC) spectra for the following three radio epochs. To collect enough counts in the MAXI spectra for a flux measurement, we combined all scans on the day of the VLA observation for epochs 4 and 5. Since the cumulative MAXI exposure differs between days, we also added all scans on the two preceding days for epoch 6. 

We used the Swift~XRT data products generator \citep{evans09}\footnote{http://www.swift.ac.uk/user\_objects/} to extract the Swift XRT spectra, while using MAXI's online on-demand tool \citep{matsuoka09}\footnote{http://maxi.riken.jp} to extra the GSC spectra. There were no nearby sources in either the automatic MAXI source or background regions. We fitted the spectra using \textsc{xspec} v12.9.0u \citep{arnaud96} with simple phenomenological models to determine the X-ray fluxes. Assuming abundances from \citet{wilms00} and cross-sections from \citet{verner96}, and using C-statistics, all spectra were well described by an absorbed power law model (\textsc{tbabs*powerlaw}). The exception is epoch 1, which required the addition of a low-temperature blackbody component (\textsc{blackbody} in \textsc{xspec}; f-test probability $\sim 2\times10^{-21}$). We did not tie the interstellar absorption parameter $N_H$ between observations as BeXRBs can show variable local absorption \citep[see e.g.][]{grinberg_cygx1}. Due to the low number of counts in the MAXI spectra (epochs 4 to 6), we fixed $N_H$ to $2 \times 10^{22}$ cm$^{-2}$ for those three epochs, additonally fixed $\Gamma$ to $1$ for epoch 4, and only attempted an absorbed power law model. Finally, we measured the unabsorbed fluxes in the $0.5-10$ keV range using the \textsc{cflux} model. All details on the spectral fits are listed in Table \ref{tab:fits}. 

For the final epoch (7), no Swift or MAXI observations were available. Therefore we used the BAT count rate instead: we assumed a similar spectral shape as during epoch 3, as this epoch also occurs during the giant outburst decay outside of the re-brightening, but also has high quality Swift XRT data. We then re-scaled the epoch 3 X-ray flux using the ratio of BAT count rates on both days. To reflect the uncertainty and assumptions in this approach, we assign a $20\%$ systematic uncertainty to the epoch 7 flux estimate. 

\section{Results}
\label{sec:results}

\begin{table*}
 \begin{center}
 \caption{\small{Observation log for our VLA monitoring of Sw J0243. The first listed epoch is part of VLA program 17B-420, while all others are part of 18A-456. The listed X-ray flux is the unabsorbed flux. Details on prior monitoring can be found in \citet{vandeneijnden18_nature}. For details on the \textit{X-ray method} and \textit{X-ray notes} columns, see Section \ref{sec:xray_methods}. All errors are quoted at $1\sigma$, while upper limits are given at $3\sigma$.}}
  \label{tab:obs}
   \begin{tabular}{lcccccccc}
  \multirow{2}{*}{} & \multirow{2}{*}{Date} & \multirow{2}{*}{MJD}         & Frequency     & Flux Density   & Spectral  & $0.5-10$ keV flux & \multirow{2}{*}{X-ray method} & \multirow{2}{*}{X-ray notes} \\
  & & & [GHz] & [$\mu$Jy] & index $\alpha$ & [erg/s/cm$^2$] & & \\ \hline\hline
  1    & 21 Feb 2018 & 58170  & 6         & $94.7 \pm 3.5$            & $-0.4 \pm 0.3$        & $(1.33 \pm 0.03)\times10^{-9}$  & Swift/XRT  & 00010467025     \\\hline
  \multirow{2}{*}{2}  & \multirow{2}{*}{09 Mar 2018} & \multirow{2}{*}{58186}  & 6         & $24.5 \pm 3.9$            & \multirow{2}{*}{$\leq -0.13$}   & \multirow{2}{*}{$(6.0 \pm 0.3)\times10^{-10}$}   & \multirow{2}{*}{Swift/XRT}  &  \multirow{2}{*}{00010467033}   \\ 
        &          &     & 22        & $< 11.5$                  &           &   &   &    \\ \hline
  3   & 27 Apr 2018 & 58235  & 4.5+7.5   & $< 14.25$                 & --        & $(7.5 \pm 0.4)\times10^{-11}$   & Swift/XRT  &  00010645011   \\ \hline
  4    & 01 May 2018 & 58239  & 4.5+7.5   & $< 15.3$                  & --        & $(5.5 \pm 1.2)\times10^{-10}$   & MAXI spectrum  & Single day    \\ \hline
  \multirow{2}{*}{5}    & \multirow{2}{*}{04 May 2018} & \multirow{2}{*}{58242}   & 4.5       & $65.0 \pm 6.0$            & \multirow{2}{*}{$-0.1 \pm 0.3$}       & \multirow{2}{*}{$(1.0 \pm 0.2)\times10^{-9}$}   & \multirow{2}{*}{MAXI spectrum}  &  \multirow{2}{*}{Single day}\\
        &         &      & 7.5       & $62.0 \pm 5.3$            &           &   &   &     \\ \hline
  \multirow{2}{*}{6}    & \multirow{2}{*}{11 May 2018} & \multirow{2}{*}{58249}   & 4.5       & $67.0 \pm 6.5$            & \multirow{2}{*}{$0.0 \pm 0.3$}       & \multirow{2}{*}{$(1.9 \pm 0.3)\times10^{-9}$}   & \multirow{2}{*}{MAXI spectrum}  & \multirow{2}{*}{Three days}    \\
        &         &      & 7.5       & $67.0 \pm 5.4$            &           &   &   &     \\ \hline
  7    & 21 May 2018 & 58259  & 4.5+7.5   & $<13.5$                   & --        & $(5 \pm 1)\times10^{-10}$   & BAT count rate  &  Using epoch 3   \\

  \hline
  \end{tabular}
  \end{center}
\end{table*}

\begin{table*}
 \begin{center}
 \caption{\small{Details of the X-ray spectral fits for VLA epochs 1 to 6 (see Section \ref{sec:xray_methods} for details on epoch 7). \textit{Observation} refers to the Swift XRT ObsID (epoch 1-3) or the MJD(s) of the MAXI scans used (epoch 4-6). The $p$ f-test refers to the comparison of a \textsc{tbabs*po} and a \textsc{tbabs*(po + bbody)} model in xspec. Parameters with an asterisk were fixed during fitting due to the low number of counts in the MAXI spectra. The low degrees of freedom (dof) for the MAXI spectra is caused by its low energy resolution. All errors are quoted at $1\sigma$.}}
  \label{tab:fits}
   \begin{tabular}{lcccccccc}
  \multirow{2}{*}{Epoch} & \multirow{2}{*}{Observation} & $N_H$ & \multirow{2}{*}{$\Gamma$} & \multirow{2}{*}{$N_{\rm po}$} & $kT_{\rm BB}$ & \multirow{2}{*}{$N_{\rm BB}$} & \multirow{2}{*}{$p$ f-test} & \multirow{2}{*}{$\chi_{\nu}^2$ (dof)} \\
  & & [$10^{22}$ cm$^{-2}$] & & & [keV] & & & \\
  \hline  \hline 
  1 & 00010467025 & $1.8 \pm 0.4$ & $2.1 \pm 0.7$ & $1.0^{+0.8}_{-0.4}\times10^{-1}$ & $2.1 \pm 0.1$ & $0.015 \pm 0.003$ & $2\times10^{-21}$ & 1.16 (754)\\
  2 & 00010467033 & $1.3 \pm 0.4$ & $0.8 \pm 0.2$ & $(3 \pm 1)\times10^{-2}$ & N/A  & N/A & $0.47$ & 0.97 (293) \\
  3 & 00010645011 & $2.1 \pm 0.4$ & $1.1 \pm 0.2$ & $(6 \pm 2)\times10^{-3}$ & N/A & N/A & $0.013$ & 1.09 (331) \\
  \hline
  4 & 58239 & $2.0^{*}$ & $1.0^{*}$& $(3.5 \pm 0.08)\times10^{-2}$ & N/A & N/A & N/A & 0.4 (6)  \\
  5 & 58242 & $2.0^{*}$ & $1.3 \pm 0.3$ & $1.0^{+0.8}_{-0.4}\times10^{-1}$& N/A & N/A & N/A & 0.4 (6) \\
  6 & 58247/48/49& $2.0^{*}$ & $1.8 \pm 0.4$ & $3.0^{+2.5}_{-1.4}\times10^{-1}$ & N/A & N/A & N/A & 1.0 (7)\\
  \hline
  \end{tabular}
  \end{center}
\end{table*}

\begin{figure*}
  \begin{center}
    \includegraphics[width=\textwidth]{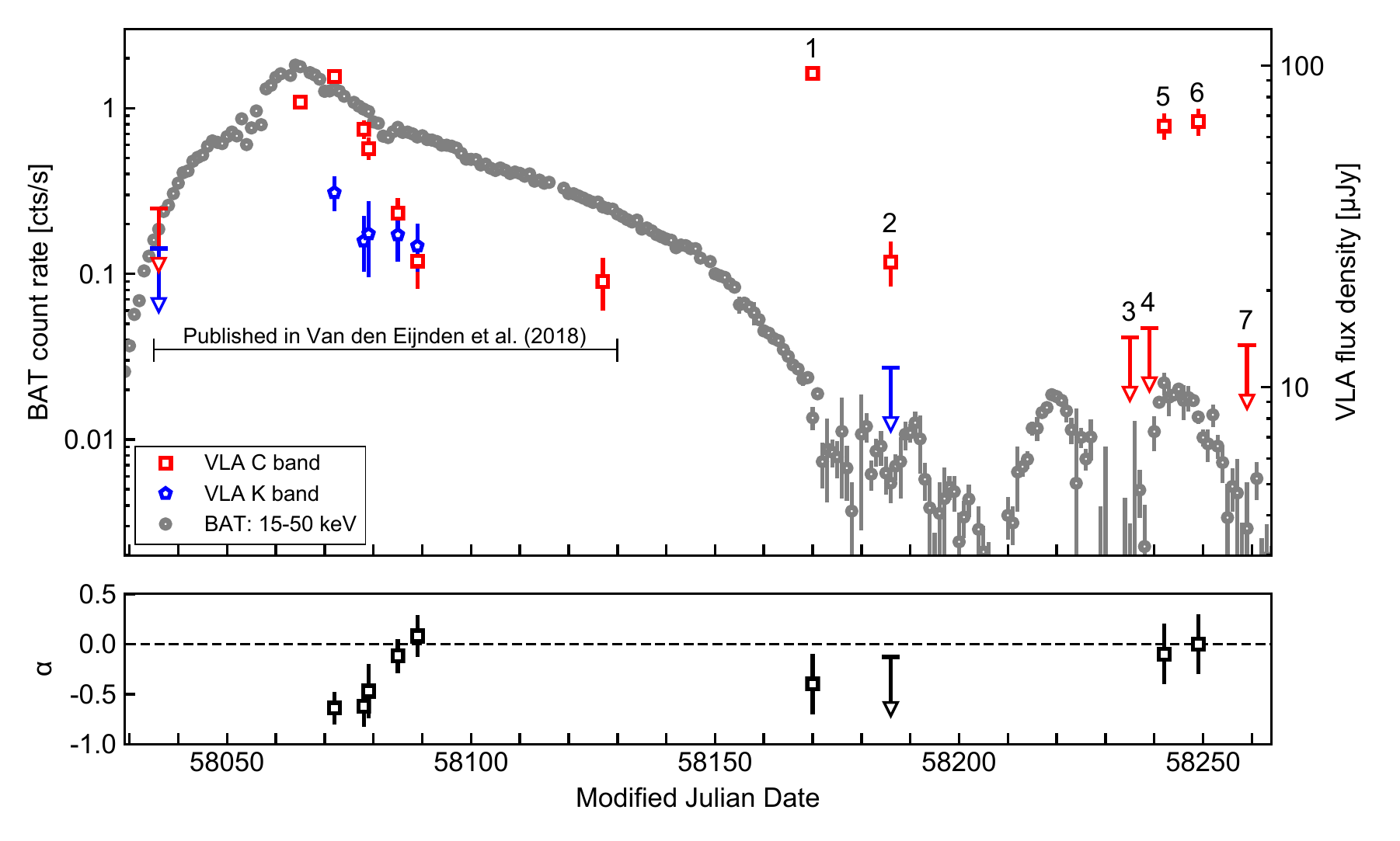}
    \caption{\textit{Top:} X-ray and radio light curves of Sw J0243's giant outburst and subsequent re-brightenings. The grey circles show the Swift BAT daily monitoring count rates, and the red squares and blue pentagons show the C and K band radio flux density, respectively. Full details about the radio flux densities and observing frequencies is listed in Table \ref{tab:obs}. This paper focuses on the last seven radio epochs, while the first eight were published previously in \citet{vandeneijnden18_nature}. \textit{Bottom:} the radio spectral index $\alpha$, where $S_{\nu} \propto \nu^\alpha$, for observations with sufficiently large frequency coverage for a measurement.}
    \label{fig:lc}
  \end{center}
\end{figure*}

\subsection{Radio and X-ray light curves}
\label{sec:results_lc}

In Fig. \ref{fig:lc}, we show the Swift BAT X-ray and VLA radio light curves of Sw J0243 since its discovery up to the end of our radio monitoring. The BAT light curve clearly reveals two re-brightenings after MJD 58200, and a possible fainter one before that date, around MJD 58190 \citep[as also visible in Fig. 2 in][]{tsygankov18}. We note that Sw J0243 has continued to show re-brightenings, of increasing duration and brightness, after the maximum plotted MJD. 

The first two (new) radio epochs (1\&2) showed an increase in radio flux, which then decayed, apparently unaccompanied by a similarly significant event in X-rays. In the first epoch, we detected the target at $6$ GHz. To estimate the radio spectrum at the time of this observation, we split the band into two halves and separately imaged each half. This provided a spectral index of $-0.4 \pm 0.3$. During the second epoch, we detected Sw J0243 at a decreased $6$ GHz flux, while not detecting the source at $22$ GHz. This non-detection also implies a negative spectral index, with a $3\sigma$ upper limit of $\alpha \leq -0.13$ (see below). The spectral shapes of both these epochs are therefore inconsistent with the steady, compact jet \citep{fender06,kording08,russell16} observed during the decay of the giant outburst \citep{vandeneijnden18_nature}. We will discuss the nature of this radio emission in Section \ref{sec:flaredisc}. 

\begin{figure}
  \begin{center}
    \includegraphics[width=\columnwidth]{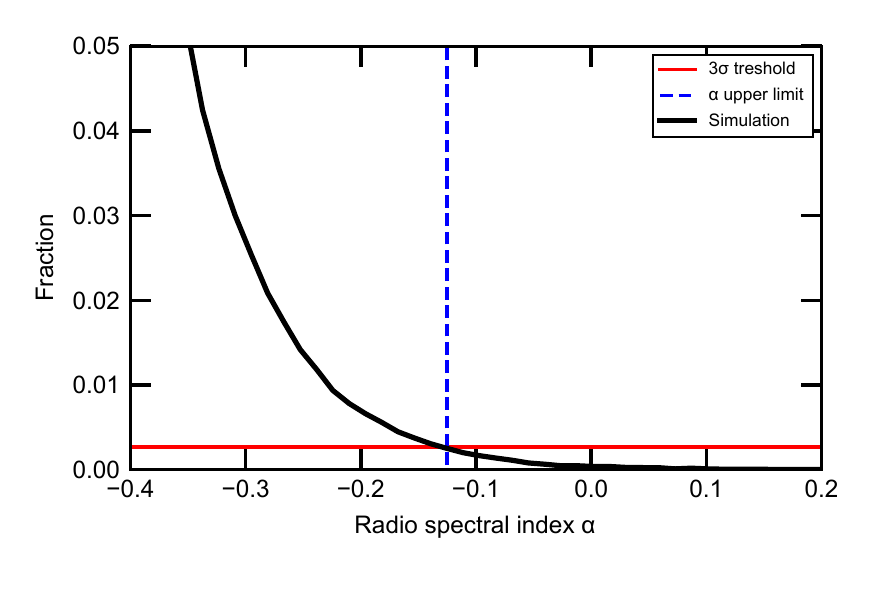}
    \caption{Monte-Carlo simulation of the spectral index $\alpha$ during epoch 2. The vertical axis denotes the fraction of simulations where, for the given spectral index, the simulated K-band flux density does not exceed three times the RMS. The upper limit on $\alpha$ is given by the location where the fraction drops below $0.27$\%.} 
    \label{fig:UL}
  \end{center}
\end{figure}

To determine the $3\sigma$ upper limit on the spectral index of epoch 2, we performed a simple Monte-Carlo simulation. We defined a grid of spectral indices between $\alpha = -0.5$ and $\alpha = 0.5$. For each value of $\alpha$, we drew $10^5$ C-band flux densities assuming a Gaussian distribution with a mean and standard deviation equalling the observed C-band flux density and RMS. For each of these simulated $6$-GHz flux densities $S(\rm 6~GHz)$, we then calculated the predicted K-band flux density as $S(\rm 22~GHz) = S(\rm 6~GHz)\times(22/6)^{\alpha}$. We then calculated the fraction of cases where the predicted 22 GHz flux density did not exceed three times the observed K-band RMS; finally, the three-sigma upper limit on $\alpha$ is obtained by determining when this fraction drops below $0.27\%$ (i.e. $100 - 99.73\%$). The results of this simulation are shown in Fig. \ref{fig:UL}.

The remaining five radio epochs (3-7) cluster around the second re-brightening. The source is only detected during the peak of this re-brightening, and at both those times the detections at $4.5$ and $7.5$ GHz show that spectrum was flat ($\alpha \approx 0.0$) The observed radio emission was surprisingly similar to the steady-jet phase of the giant outburst \citep{vandeneijnden18_nature}. Combined with the clear connection to the X-ray flux increase, this similarity strongly implies we observed a jet during the re-brightening as well. 

The radio emission brightens and disappears remarkably abruptly (on time-scales of days), with similarly deep non-detections before and after the peak of the re-brightening. In fact, the X-ray flux measurements during epochs 3 and 4 indicate that the re-brightening had already started several days earlier, increasing in X-rays by at least a factor $7.3$ without any radio emission observed in either epoch. The jet then established within three days between epochs 4 and 5, increasing at least an order of magnitude in radio luminosity despite only a minor increase in X-ray flux (less than a factor two). As the X-ray flux subsequently dropped in epoch 7, back to the level of epoch 4, the jet disappeared similarly quickly.  

\subsection{The X-ray -- radio luminosity plane}

\begin{figure*}
  \begin{center}
    \includegraphics[width=\textwidth]{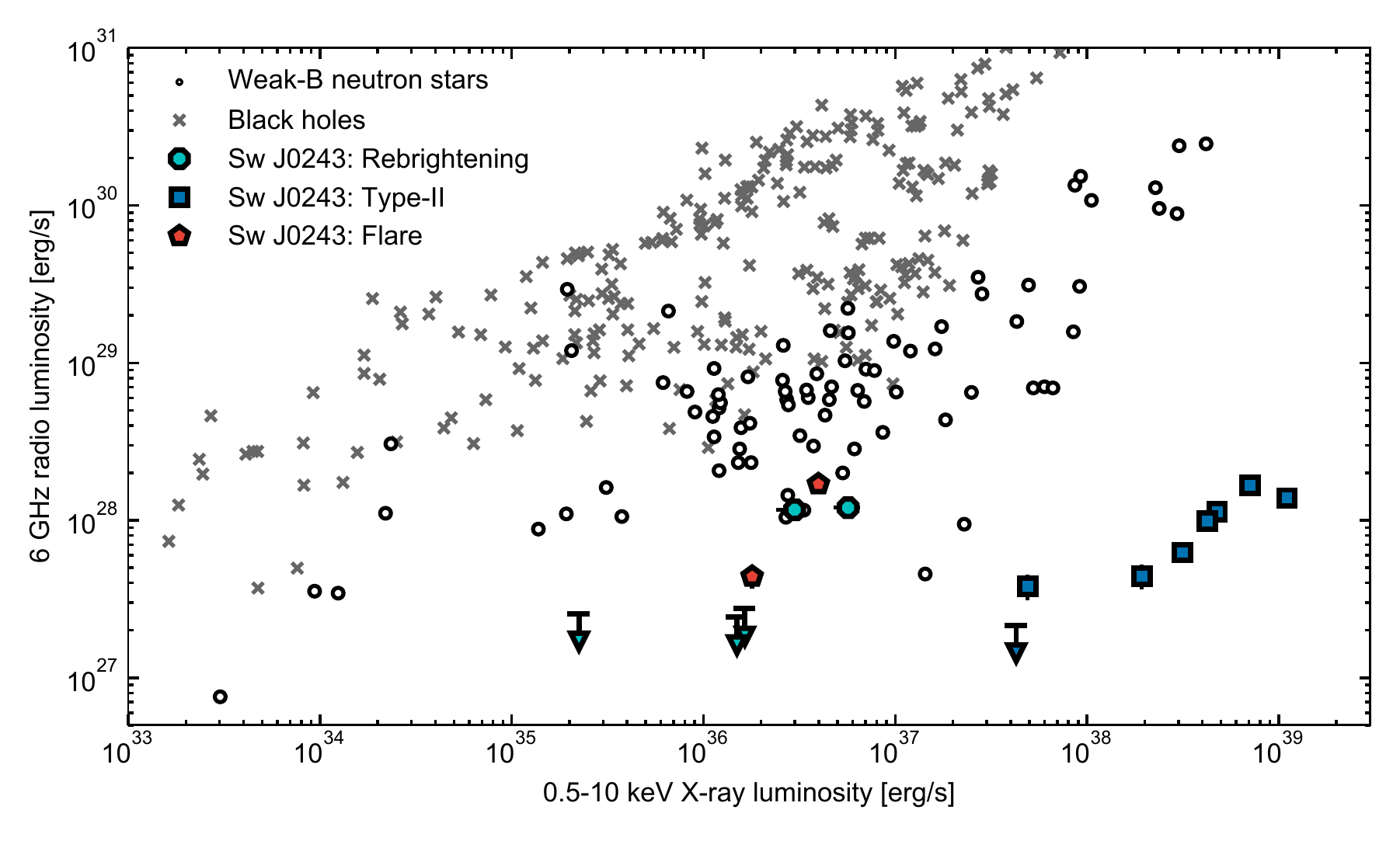}
    \caption{The X-ray/radio luminosity plane for X-ray binaries. We plot a sample of both BHs and weakly-magnetized NSs in LMXBs \citep[from][]{gallo18}. Additionally, we plot our monitoring observations of Sw J0243, labeled based on the stage in the outburst (giant outburst, flare, re-brightening) and assuming a distance of $5$ kpc \citep{vandeneijnden18_nature}. We follow common practice and rescale all luminosities to $6$ GHz assuming a flat spectrum. Remarkably, the Sw J0243 jet is as radio bright during both the giant outburst and the re-brightening. Additionally, it switches on suddenly during the re-brightening as $L_X$ passes $\sim 2\times10^{36}$ erg/s.} 
    \label{fig:lxlr}
  \end{center}
\end{figure*}

In Fig. \ref{fig:lxlr}, we show the X-ray -- radio luminosity plane for X-ray binaries, plotting all radio observations of Sw J0243 alongside a sample of accreting BHs and NS LMXBs \citep[collected by][]{gallo18}\footnote{The plotted sample is publicly available at \href{https://github.com/jvandeneijnden/XRB-Lx-Lr-Sample}{https://github.com/jvandeneijnden/XRB-Lx-Lr-Sample}}. While during the giant outburst the jet was orders of magnitude fainter than those seen in NS LMXBs accreting at similar X-ray luminosity (e.g. the Z-sources), it is consistent with the lower end of the NS population during epoch 1 and the radio re-brightenings (epochs 5 and 6). While it is remarkable that epoch 1 appears to be similar in radio brightness to both the giant outburst peak and the re-brightening, due to the sparse sampling, there is a considerable chance that this epoch does not coincide with the maximum flux around this time in the outburst. 

Additional interesting comparisons can be made in the horizontal direction -- between the giant outburst and the re-brightening -- and the vertical direction -- between the detections and non-detections during the re-brightening. The latter highlights the point made in Section \ref{sec:results_lc}: while the radio luminosity does not exceed the radio detection limit between X-ray luminosities of $\sim 2\times10^{35}$ and $\sim 1.5\times10^{36}$ erg/s, the radio jet swiftly appears a factor $\sim 5$ above the radio non-detection as $L_X$ passes $\sim 2\times10^{36}$ erg/s \citep[all assuming a distance of 5 kpc;][]{vandeneijnden18_nature}. Similarly, the jet, if still present, returns to non-detectable flux levels quickly as the X-ray luminosity decays below this threshold. 

The horizontal comparison between the peaks of the re-brightening and giant outburst reveals a similar maximum radio luminosity, despite more than two orders of magnitude difference in X-ray luminosity. Given the sampling of our radio monitoring of the giant outburst \citep[i.e.][]{vandeneijnden18_nature}, it is likely we measured the peak radio flux density during that stage. The sampling during the re-brightening is more sparse, so the peak radio luminosity might be higher than what we measure; even in that case, the similarity in the detected radio flux density levels despite the orders of magnitude lower mass accretion rate is striking. 

\section{Discussion}
\label{sec:discussion}

With continued VLA radio monitoring of the BeXRB Sw J0243, we have observed radio emission during the late decay of the giant outburst and, for the first time, detected a BeXRB in radio during an X-ray re-brightening. Most remarkably, the latter radio emission is as bright as during the super-Eddington phase of the giant outburst, and turns on and off swiftly (on time-scales of days) as the X-ray luminosity passes $2\times10^{36} (D/5 \rm kpc)^2$ erg/s. Here, we will discuss the radio flaring and X-ray re-brightening separately, focussing on both the origin of the radio jet and the accretion process, by combining for the first time X-ray and radio monitoring of a NS BeXRB. We will separate the first two (1 \& 2) and final five (3 to 7) epochs presented in this paper, given the difference in outburst stage and the substantial separation in time. 

\subsection{Epochs 1 and 2: a fading radio flare}
\label{sec:flaredisc}

Epochs 1 and 2, during the late decay of the giant outburst, show varying radio flux densities. Epoch 1 shows bright radio emission, higher by a factor $\sim 4$ than during the preceding epoch and epoch 2, which show similar flux densities. During both epochs 1 and 2 we observe a steep ($\alpha < 0$) spectrum, which is inconsistent with a compact jet \citep{fender06} which was seen during the decay from the super-Eddington peak and during the re-brightening (Section \ref{sec:turnon_off}). The similarity in spectral shape between epoch 1 and 2 suggests that the emission originates from the same process, for instance a decaying radio flare. Alternatively, since no radio observation was taken for $\sim 50$ days before epoch 1, another possibility is that only epoch 1 shows radio flaring while the preceding epoch and epoch 2 represent a contant base level of radio emission. However, that does not explain the similar radio spectral shape in epoch 1 and 2, and would require that the radio emission remains constant over months while the X-rays continually decay. 

In both scenarios, at least epoch 1 shows radio flaring of Sw J0243. Different mechanisms could cause such flaring radio emission. For instance, the increased flux density and spectral shape might be caused by the launch of discrete ejecta \citep{fender06}. However, around epoch 1, the BAT X-ray light curve shows a stable, gradual decay without any significant X-ray event. As visible in Fig. \ref{fig:lc}, there are hints of a first, weak re-brightening starting around MJD 58190. However, this possible re-brightening occurs just after epoch 2 and therefore cannot be causally connected to the launch of discrete ejecta. Based on this lack of a clear X-ray event, we deem this first explanation for epochs 1 and 2 unlikely. For more considerations about radio flaring without an accompanying X-ray event in a NS X-ray binary, we refer the reader to the extensive discussion in \citet{tudor17}.

Further, an origin in the jet itself, unrelated to the accretion flow, could be the interaction of outflowing material with either the remnants of the pre-existing jet, or the ISM. Shocks in such a scenario could produce the optically thin specrum observed during epoch 2. However, as Sw J0243's outflows remain unresolved at the resolution of the VLA, we cannot see such interactions directly. In this scenario, the similarity of the epoch 1 flux density to the radio peak during both the giant outburst and the re-brightening epochs (i.e. 3 to 7) is a coincidence; the alternative is unlikely given our sparse sampling during the flare, reducing the chances that we actually observed the peak of the flare. 

\subsection{The quickly re-establishing jet during re-brightenings}
\label{sec:turnon_off}

The detected radio emission during the X-ray re-brightening shows clear similarities to the jet during the giant outburst: we observe flat spectrum radio emission, consistent with a steady, compact outflow \citep{fender06,kording08,russell13,russell16}, with the luminosity levels consistent with those of the earlier jet. Furthermore, the radio emission is directly linked to the X-ray luminosity, as we will discuss in detail below. As argued by \citet{vandeneijnden18_nature} for the radio detections obtained during the giant outburst, alternative radio emission processes cannot account for the combination of observed radio properties during the re-brightening: the stellar wind is expected to be orders of magnitude fainter \citep[e.g][]{wright75,kording08,russell16}, the onset of a radio emitting outflow only at the highest X-ray luminosities is the opposite behaviour to that of a magnetic propeller \citep{illarionov75}, and a flat spectrum is not expected for a shock origin. Therefore, we conclude that the radio emission during the X-ray re-brightenings also appears to originate from a compact jet. 

During the re-brightening, the jet only reaches detectable radio luminosities close in time to the X-ray peak. Between the radio non-detections in epochs 3 and 4, the accretion flow brightens by almost an order of magnitude in four days, while during the three days between the epoch 4 and the first re-brightening radio detection (epoch 5), the X-ray luminosity only increases by a factor less than 2. During the decay of the re-brightening, the radio jet has again turned off around the same apparent X-ray threshold of $2\times10^{36} (D/5 \rm kpc)^2$. The $\sim$~days time scale of the turn on and off of the jet is remarkably shorter than observed for jets in other X-ray binaries, where this process (and large scale evolution of the jet in general) typically takes one to a few weeks \citep[e.g][]{corbel13,vanderhorst13,russell14}. In addition, \citet{tsygankov18} show that while the X-ray spectral shape gradually evolves with X-ray luminosity, no dramatic spectral change occurs at the above X-ray luminosity threshold. 

This begs the questions: how can the jet establish so rapidly, and what is changing in the accretion flow and jet formation process around an X-ray luminosity of $2\times10^{36} (D/5 \rm kpc)^2$? Although this behaviour is currently only observed in Sw J0243, and remains to be confirmed in other BeXRBs (see Section \ref{sec:and_what_now}), we will discuss one possible explanation invoking the interaction between the NS magnetic field and the accretion flow below. 

\subsubsection{Magnetic accretion onto neutron stars}
\label{sec:model}

When considering accretion onto a (strongly) magnetized object, the presence of a single X-ray luminosity threshold between distinct accretion and/or outflow states, naturally brings to mind the interaction between the magnetic field and the accretion flow. In particular, the accretion process can be divided into two states, depending on the location of the magnetospheric radius $R_{\rm m}$ (where the accretion disk ram pressure and the magnetic field pressure are equal): $R_{\rm m}$ is located either inside or outside the co-rotation radius $R_{\rm co}$, where the accretion flow orbital and NS spin frequencies are equal \citep[see e.g.][]{illarionov75,cackett09,dangelo10,dangelo12,tsygankov16a,tsygankov17}. When $R_{\rm m} > R_{\rm co}$, which occurs at relatively low accretion rates, the rotating magnetic field creates a centrifugal barrier for the inflowing material. Outflows might be launched in this state per the propeller mechanism. Conversely, when the magnetospheric radius is pushed inwards at higher mass accretion rates and $R_{\rm m} < R_{\rm co}$, matter can instead flow in undisturbed by the magnetic field. These states are separated by the case where both radii are equal, which occurs at a single X-ray luminosity referred to as the limiting luminosity or $L_{\rm lim}$. 

The importance of the magnetospheric and corotation radii in BeXRB accretion flows was noted decades ago in order to explain the correlation between orbital period and spin period in these sources \citep{davidson73,corbet85,corbet86,corbet87,stella86,waters89}. As discussed in more detail in Section \ref{subsec:pulsations}, the NS can either be spun up or down due to the angular momentum difference with the accretion flow. To first order, this difference in angular momentum depends on the relative positions of the magnetospheric radius and the corotation radius \citep{ghoshlamb}. As BeXRB spin periods are relatively stable over long time-scales (e.g. covering multiple outbursts), the spin up during outburst and spin down during quiescence should approximately cancel. The relation between the equilibrium spin period and orbital period can then be explained by the dependency of the mass accretion rate, which affects the magnetospheric radius, on the orbital period. As a BeXRB goes into outburst, it should therefore cross the limiting luminosity where $R_{\rm m}$ exactly equals $R_{\rm co}$.\footnote{We note that systems with very slow spins, exceeding $\sim 100$ seconds, might show a \textit{cold disk} state, where the ionization fraction of the accretion flow is very low. In that scenario, the interaction between the disk and NS magnetic field might be different \citep{tsygankovwijnands17,tsygankov17}.}

Although it remains unclear where exactly a jet is launched when the NS magnetic field exceeds $\sim 10^9$ G \citep[see e.g.][]{massi08,vandeneijnden18_nature}, we can naively expect that the jet launching region is located inside the corotation radius -- approximately $3.7 \times 10^3 R_g$ for Sw J0243, assuming a $1.4$ $M_{\odot}$ neutron star. A magnetic centrifugal barrier might then prevent material from (efficiently) reaching this jet launching region. That implies that a jet could only be formed when $R_{\rm m} < R_{\rm co}$ or, equivalently, $L_X > L_{\rm lim}$. Such a scenario naturally provides a single X-ray luminosity for a specific source where the jet would turn on and off. In the remainder of Section \ref{sec:turnon_off}, we will explore this idea and its implications further. 

Before we do so, it is important to note two caveats: firstly, jet formation is not understood for strongly magnetized neutron stars (see also Section \ref{sec:how_does_the_jet_work}) and we assume, a priori, that the jet originates from within the corotation radius. Secondly, we also note that the radio non-detection at the start of the giant outburst occured when $L_X > L_{\rm lim}$. However, X-ray binaries are known to show different behaviour during the outburst rise and decay \citep[e.g][]{fender04,gilfanov10,munozdarias14} and currently no observational constraints exist on the coupling between the jet and accretion flow during the outburst rise of BeXRBs. Therefore we cannot conclude whether during the early rise no jet is launched, or it is simply too faint to detect.

\subsubsection{The magnetic field of Sw J0243}
\label{sec:magfield}

In the above scenario for the jets during the re-brightening of Sw J0243, the transition X-ray luminosity of $2\times10^{36} (D/5 \rm kpc)^2$ erg/s would equal $L_{\rm lim}$. The limiting X-ray luminosity can be estimated as \citep{tsygankov17}:
\begin{equation}
    L_{\rm lim} = 4\times10^{37} k^{7/2} B_{12}^{2} P^{-7/3} M_{1.4}^{-2/3} R_6^5 \text{ erg s}^{-1}
    \label{eq:Llim}
\end{equation}
\noindent where k is a geometric correction factor, typically assumed to be $1/2$ for disk and $1$ for spherical accretion \citep{ghoshlamb}, $B_{12}$ is the NS magnetic field strength in units of $10^{12}$ G, P is the NS spin period in seconds, $M_{1.4}$ is the NS mass in terms of $1.4 M_{\odot}$, and $R_6$ is the NS radius in $10^6$ cm. Setting $L_{\rm lim} = 2\times10^{36}$ erg/s, assuming canonical NS parameters and disk accretion, we estimate $B \approx 10^{13}$ G. Taking into account that the Gaia DR2 distance of $5$ kpc is merely a lower limit \citep{vandeneijnden18_nature}, $L_{\rm lim}$ gives a magnetic field of $B \gtrsim 10^{13}$ G. 

With the lack of any observed cyclotron features in the X-ray spectrum of Sw J0243, its magnetic field strength has not been measured directly. Indirect, model-dependent estimates have been made by searching for the onset of a magnetic propeller \citep{tsygankov18} and the modelling of the X-ray pulse profiles and frequency evolution \citep{tsygankov18,wilson18}. The lack of an observed onset of a propeller implies a magnetic field lower than $10^{13}$~G (however, see below). Using the pulse profiles and evolution implies a magnetic field exceeding $10^{13}$~G, both using only Fermi/GBM and using Fermi/GBM + NICER monitoring. While we stress that these estimates are model dependent and not direct measurements, they do support that the required magnetic field for $L_{\rm lim} = 2\times10^{36}$ erg/s is plausible for Sw J0243. 

As discussed extensively in \citet{tsygankov18}, Sw J0243 is not observed to transition to the propeller regime \citep{illarionov75}, wherein material is expelled by the centrifugal barrier of the rotating magnetic field when $R_{\rm m} > R_{\rm co}$ (i.e. $L_X < L_{\rm lim}$). Such a propeller is expected to be accompanied by a sudden decrease in X-ray flux and possibly by brightened and erratically varying radio luminosities \citep{tudor17}, which are both not observed. However, the presence of a centrifugal magnetic barrier does not necessarily lead to mass loss via an outflow from the system; as noted by \citet{sunyaev77}, \citet{spruit93}, and \citet{dangelo10,dangelo11,dangelo12}, there are well-defined solutions to the interaction of a spinning magnetosphere inside a thin accretion disk which do not invoke mass loss through an outflow. In those scenarios, matter accumulates outside the magnetospheric radius instead of being expelled. If such a scenario is at play in Sw J0243, it could explain the lack of observed propeller characteristics despite a NS magnetic field exceeding $10^{13}$~G. 

\subsubsection{Pulse frequency evolution}
\label{subsec:pulsations}
The location of the magnetospheric radius with respect to the corotation radius can influence the evolution of the spin frequency of the NS: at high mass-accretion rate, when $R_{\rm m} < R_{\rm co}$, the accreted material has a higher angular momentum than the NS. In the opposite case, the NS can instead lose angular momentum to the accretion flow instead \citep{ghoshlamb,alpar82,Radhakrishnan82}. Without any other processes affecting the NS spin, one could therefore predict that, to first order, the sign of the NS spin derivative should change when the magnetospheric and corotation radius are exactly equal. This picture is quite simplistic and the predicted behaviour is often not observed in long-term monitoring of accreting NSs \citep[e.g.][]{bildsten97,chatterjee00,patruno10b}. However, we can still explore the spin frequency evolution for Sw J0243, especially during the re-brightening targeted with the VLA. 

The pulsation frequency of Sw J0243 has been monitored by Fermi GBM and NICER throughout the outburst. Due to the orbital motion of the NS, the observed frequencies have to be corrected by the ephemeris to study the intrinsic spin evolution. Therefore, (slight) differences exist between the three available spin histories of Sw J0243: a combined GBM + NICER analysis presented by \citet{wilson18}, a GBM-only analysis by \citet{doroshenko18}, and finally the entire GBM monitoring available on the GBM Accreting Pulsar Histories webpage\footnote{\href{https://gammaray.nsstc.nasa.gov/gbm/science/pulsars.html}{https://gammaray.nsstc.nasa.gov/gbm/science/pulsars.html}}. Broadly speaking, all show that Sw J0243 mainly spun up during the giant outburst, but might have turned to a spin down phase afterwards during the late outburst decay. 

Only the GBM Accreting Pulsar Histories webpage currently contains spin measurements during the re-brightening covered by the VLA, as both others cut off earlier in time. However, at the time of writing, these measurements during the re-brightening are corrected with the ephemeris in \citet{jenke18}, which differs significantly from the more up to date ephemeris derived by \citet{wilson18}; indeed, on the GBM monitoring webpage, residual variations in the spin frequencies can be seen on the orbital time scale during the re-brightenings. 

Given the known deviations of accreting NSs from the simple accretion-induced spin evolution scenario presented above, and these challenges with currently available monitoring, we cannot make detailed inferences for Sw J0243. However, such spin evolution monitoring, especially of known sources with well-constrained ephemeris, can be a valuable addition to future radio observations of BeXRB re-brightenings and help constrain our interpretation in Sections \ref{sec:model} and \ref{sec:magfield}. 

\subsubsection{Comparison to AMXP jet formation}

The proposed requirement that the magnetospheric radius is located within the corotation radius for jet launching to be possible, can of course be tested for other sources as well. With their much faster spin and jet detections at lower X-ray luminosities, the accreting milli-second X-ray pulsars, AMXPs, are especially interesting in this regard: the faster spin decreases the corotation radius, while at lower X-ray luminosity the magnetospheric radius is larger.

The two AMXPs with radio detections at the lowest X-ray luminosities are SAX J1808.4-3658, at $\sim 2 \times 10^{34}$ erg/s, and IGR J00291+5934, at $\sim 1\times10^{34}$ erg/s \citep{tudor17}. These NSs rotate at $401$ Hz \citep{wijnands98} and $599$ Hz \citep{galloway05}, respectively. For both sources, assuming that a jet detection can only occur when $L_X \geq L_{\rm lim}$ (e.g. Eq. \ref{eq:Llim}), we can derive an upper limit on the magnetic field strength. Assuming canonical NS parameters and disk accretion (i.e. $k=0.5$), the measured NS spin and the lowest X-ray luminosity where a jet is still detected, we infer $B \leq 7\times10^7$ G and $B \leq 3\times10^7$ G for SAX J1808.4-3658 and IGR J00291+5934, respectively. For both sources, these rough estimates are relatively low, but still broadly consistent with earlier estimates based on the truncation of the accretion flow by the magnetic field \citep{mukherjee15}. 

For both AMXPs (or any other AMXP), the magnetic field has not been measured directly but only estimated, typically through the interaction between the accretion flow and the magnetic field. If the two AMXPs discussed above in reality have stronger magnetic fields than our above upper limits, this would imply either (i) that our argument that $L_X \geq L_{\rm lim}$ for jet formation is invalid, or (ii) that NSs with different magnetic field strengths and spin periods can launch jets through different mechanisms. Given the gaps in the current understanding of NS jets, especially for strong magnetic fields, the second scenario cannot currently be discarded. 

\subsection{The origin of the X-ray re-brightenings}
\label{sec:re-brightenings_origin}

Our scenario for the sudden re-establishment of the radio jet leaves two vital questions unaddressed: what drives the changes in mass accretion rate responsible for the X-ray re-brightenings and moving the magnetospheric radius in- and outwards, and why should material move inside the corotation radius for jet formation to be possible? We will discuss the former question in this section, and focus on the formation of the jet in Section \ref{sec:how_does_the_jet_work}. 

\hyphenation{Wij-nands}
X-ray re-brightenings during outburst decays are seen in dwarf novae and occasionally in LMXBs \citep[see for an overview][]{chen97}, such as the NSs SAX J1808.4-3658 \citep{patruno09,sanna15,patruno16} and IGR J17379-3747 \citep{vandeneijnden18_17379} and the BHs XTE J1650-500 \citep{tomsick04}, Swift J1910.2-0546 \citep{tomsick13}, GRS 1739-278 \citep{yan17}, and MAXI J1535-571 \citep{parikh18}. As discussed in detail in \citet{patruno16}, these re-brightenings appear to be caused by the hydrogen-ionisation instability, which is also responsible for the main outburst \citep{lasota01}. It remains harder to explain why the instability is triggered at the end of the main outburst, although this behaviour is also seen to occur spontaneously in simulations by \citet{dubus01}, or might be related to increased irradiation by the donor \citep{hameury00}. 

X-ray re-brightenings are often observed in BeXRBs as well \citep[e.g.][]{chen97,sguera17,rouco17}, although their origin remains debated; some recent examples include 4U 0115+63 and V 0332+53 as discussed by \citet{tsygankov16b} and \citet{wijnands16}, or the recent giant outbursts of XTE J1946+274 and KS 1947+300\footnote{See e.g. the ESA Be/X-ray binary monitor; \href{http://integral.esac.esa.int/bexrbmonitor/webpage_oneplot.php}{http://integral.esac.esa.int/bexrbmonitor/webpage\_oneplot.php}}. The above explanation of the re-brightenings in LMXBs has also been proposed for BeXRBs \citep[e.g.][]{tsygankovwijnands17}, and could be at play in Sw J0243. The time-scale, from weeks to months, of the re-brightenings in Sw J0243 appear feasible for such a scenario; these are similar to the durations of outbursts of, for instance, faint transient NS LMXBs, which can be explained by the hydrogen-ionization instability model for X-ray binary outbursts \citep{lasota01}. Interestingly, \citet{patruno16} argue for the NS LMXB SAX J1808.4-3658 that the re-brightenings only occur when $R_{\rm m} \sim R_{\rm co}$, which fits with our interpretation of the radio behaviour of Sw J0243 (Section \ref{sec:turnon_off}). 

Additionally, for BeXRBs the re-brightenings could be connected to Type-I outbursts, which occur at periastron as the NS moves through the Be-companion's circumstellar disk. The re-brightenings are not always observed at periastron in every source (e.g. in XTE J1946+274), and hence this connection may not explain all re-brightenings. However, we can also consider this scenario for Sw J0243. Since its discovery, four publications have reported measurements of the ephemeris based on pulsation modeling \citep{ge17, jenke18, doroshenko18, wilson18}. These all use the Fermi/GBM pulse monitoring, with \citet{ge17} adding HMXT/Insight observations and \citet{wilson18} including NICER monitoring. The epochs of periastron between these four measurements differ up to three days, and the orbital periods by approximately one day. Therefore, it is difficult to unambiguously relate the re-brightening peaks in Sw J0243 with periastron passages. 

At the time of writing, the re-brightenings are also continuing and increasing in duration, lasting longer than the orbital period\footnote{See e.g. \href{https://swift.gsfc.nasa.gov/results/transients/weak/SwiftJ0243.6p6124/}{https://swift.gsfc.nasa.gov/results/transients/weak/} \href{https://swift.gsfc.nasa.gov/results/transients/weak/SwiftJ0243.6p6124/}{SwiftJ0243.6p6124/}}. In contrast, Type-I outbursts typically last up to $\sim 30\%$ of an orbit \citep{reig11}. However, the presence of a residual accretion flow from the preceding giant outburst might change the effect of periastron passage in the BeXRBs; for instance, an increase in mass-accretion rate at periastron could trigger the hydrogen-ionization instability \citep[e.g.][]{lasota01} in the remaining material, creating a longer and brighter outburst than during isolated periastron passages. In such a scenario, it is merely the trigger of the instability that differs from LMXBs. 

Finally, our earlier arguments about the magnetospheric and corotation radii bring to mind the presence of a trapped disk \citep{sunyaev77,dangelo10}. In that scenario, a reservoir of material builds up when the centrifugal barrier is present, gradually increasing the disk ram pressure and moving in the magnetospheric radius. Eventually this will cause the centrifugal barrier to disappear and accretion to ensue undisrupted. While an interesting option in the context of the sudden switching on of the jets, the time-scales of the trapped disk do not work out for the X-ray evolution. Assuming an accretion disk efficiency of $\sim 10\%$ and the simulations of this state in \citet{dangelo12}, we find time-scales of $1-10\%$ of the viscous time-scale at the corotation radius. Despite the large corotation radius of Sw J0243, the $\sim 1$ day viscous time-scale at this radius \citep[e.g.][]{frank92} implies that the time-scale of the trapped disk is not reconcilable with the $\sim $ month time-scale of the re-brightenings.  

\subsection{Jets from strongly-magnetized neutron stars}
\label{sec:how_does_the_jet_work}

In Section \ref{sec:turnon_off}, we assumed that material needs to pass the corotation radius for a jet to be formed. This assumption is not based in jet theory; in fact, jets from strongly magnetized NSs ($B \geq 10^{12}$ G) currently are poorly understood theoretically. While jets are ubiquitous in accreting, fast-spinning NSs with weak magnetic fields below $\sim 10^9$ G -- none of which reside in HMXBs -- the jet observed during the giant outburst of Sw J0243 was the first clear jet detection from a strongly-magnetized NS. This first detection was preceded by decades of jet non-detections in both large surveys \citep[i.e.][]{duldig79,nelson88,fender00,migliari06} and observations of individual targets \citep{tudose10, migliari11}. Recently, radio emission was detected from the two strongly-magnetized accreting NSs GX 1+4 \citep{vandeneijnden18_gx} and Her X-1 \citep{vandeneijnden18_her}. However, the exact origin of this radio emission remains undetermined as limited spectral and temporal information made a direct jet inference impossible. 

Classical \textit{magneto-centrifugal} jet models, most commonly assumed for accreting NSs (or any object without an event horizon) predict that a strongly-magnetized NS cannot launch a jet \citep{blandford82, massi08}. Therefore, the jet observed in Sw J0243 requires a different explanation. One such option is the model introduced by \citet{parfrey16}, where the jet power is not provided by the accretion disk rotation, but instead by NS magnetic field lines opened up by the accretion flow. Interestingly, this model correctly predicts the faintness of the jet in Sw J0243 compared to the Z-sources, based on the stronger magnetic field and slower spin of the former. 

However, our jet monitoring during the re-brightening complicates the picture. During the re-brightening, the jet becomes as radio bright as it was during the super-Eddington giant outburst peak, despite over two orders of magnitude difference in X-ray luminosity. While the jet is remarkably faint during the giant outburst, it is consistent with the lower end of the NS LMXB radio luminosities during the re-brightening (see Fig. \ref{fig:lxlr}). This consistency implies that the NS parameters do not regulate jet power in such a simple way as predicted in the \citet{parfrey16} model, assuming of course that a single formation mechanism underlies the jet in both the super- and sub-Eddington states. The future detectability of jets from strongly-magnetized NSs is greatly increased by this more complex relation; if these jets were as suppressed compared to weakly-magnetized NSs around $10^{36}$--$10^{37}$ erg/s as during the super-Eddington stage, current generation radio telescopes could not detect them during the fainter states. 

Assuming a single jet launching mechanism during both the giant outburst and the re-brightenings is not necessarily valid. Given the significant differences in accretion flow properties, the possibility exists that the two jets are launched through different mechanisms and reach similar maximum radio luminosities by coincidence. In the \citet{parfrey16} jet model, the power to expel material from the disk is provided by the accretion flow opening up magnetic field lines; the rotation of the disk then collimates the outflow into a jet. The main requirement for this mechanism is that the accretion flow penetrates into the light cylinder radius of the NS ($> 1.4\times10^6$ $R_g$ for Sw J0243). Our proposed requirement for jet formation during the re-brightenings -- $R_{\rm m} < R_{\rm co}$ -- is much stricter and does not appear in the \citet{parfrey16} model. Therefore, it might be possible that different jet launching mechanisms are at play in Sw J0243, with the mechanism by \citet{parfrey16} only becoming dominant at the highest X-ray luminosities. However, the currently available observational constraints from Sw J0243 are not sufficient to further consider such a scenario. 

\subsubsection{What do the X-ray and radio luminosity trace?}

When comparing the radio emission during the giant outburst and the re-brightening, one can ask what the X-ray and radio luminosities trace: how can the observed radio flux densities be so similar at such different mass accretion rates? Or, alternatively, do the measured radio luminosities actually probe the jet power in a similar fashion? Below, we explore these questions, assuming for simplicity a single jet launching mechanism during both accretion regimes. 

The accretion flow geometry and structure during the decay and re-brightenings of BeXRB giant outbursts is not fully understood \citep[][see Section \ref{sec:re-brightenings_origin}]{reig11}. Similar uncertainty currently exists about the properties of super-Eddington accretion flows onto NSs, as seen during the peak of some giant outbursts of BeXRBs and in Ultra-Luminous X-ray pulsars (ULXPs). However, it is expected that significant differences between the accretion flow in these two states exist; for instance, radiation pressure at super-Eddington mass accretion rates is thought to puff up the inner accretion flow into a vertically extended (i.e. thick) accretion flow in ULXPs \citep[e.g.][]{mushtukov17,walton18}, compared to a thin disk at lower accretion rates. However, directly comparing the X-ray luminosity of these two states might not trace the same physical structures in the X-ray binary. In other words, not all X-ray flux is necessarily related to the launch of the jet. 

While it is unclear where exactly in NS X-ray binaries the jets are launched, all characteristic radii in the accretion flow (i.e. $R_{\rm m}$, $R_{\rm co}$) are further out for strongly-magnetized, slowly spinning NSs than for their millisecond, weakly-magnetized counterparts. Therefore, one could naively assume that, if it exists, any radius characteristic of the jet launching (i.e. where matter is funneled into the outflow), moves out with increasing magnetic field strength and decreasing spin. As stated before, the accretion flow changes significantly in structure between the super-Eddington regime of the giant outburst, and the re-brightening at much lower $L_X$. However, if these significant changes predominantly occur within the -- relatively far out -- characteristic jet launching radius, the X-ray luminosity effectively coupled to the jet might not increase as much. In other words, could the jet be coupled only to the outer accretion flow (and its related X-ray luminosity), while the difference in X-ray luminosity between the super- and sub-Eddington phases is dominated by the inner flow?

There are several issues with this simplistic picture. Firstly, the mass-accretion rate in the inner flow cannot increase to super-Eddington values, without the mass accretion rate throughout the entire disk increasing. From an observational point of view, this scenario is at odds with the coupling between the X-ray and radio luminosities observed during the giant outburst; the jet would instead be expected to remain at roughly constant power. Finally, comparing the X-ray spectra between the giant outburst \citep{vandeneijnden18_nature} and the re-brightening does not reveal large systematic changes signalling a completely different inner accretion flow structure dominating the flux during the former. 

In all of this, we make the common assumption that the radio luminosity at a single frequency -- in this case $6$ GHz -- is representative of jet power in all types of X-ray binaries \citep[see][for more issues with this assumption]{russell14}. This assumption can however lead to apparent differences between observations, that do not necessarily exist; for instance, \citet{espinasse18} recently demonstrated that the differences between the radio quiet and loud tracks of BHs in the radio/X-ray luminosity can be affected by the radio spectral shape. In similar fashion, the comparable radio luminosities of Sw J0243 might not map onto the same jet power. 

For instance, there might be a difference in break frequency of the jet spectrum, with the jet spectrum extending to higher frequencies at higher mass accretion rates. Such changes in jet break frequency have been observed both in BHs \citep[e.g.][]{russell14} and NSs \citep[e.g.][]{diaztrigo18}, although their evolution with accretion flow geometry remain unclear. In that case, simply comparing the $6$ GHz luminosity would disregard a large fraction of jet power during the giant outburst. For instance, there might be a difference in break frequency or an evolution of the synchrotron cooling break, both of which significantly impact the broadband jet spectrum at different mass accretion rates. Changes in jet break frequency have been observed both in BHs \citep[e.g.][]{russell14} and NSs \citep[e.g.][]{diaztrigo18}, although their evolution with accretion flow geometry remain unclear. Additionally, while the location and behaviour of the high-energy synchrotron cooling break is not well understood \citep{peer12,shahbaz13,gardner13}, its position and evolution will dictate the total radiative jet power \citep[][]{russell14}. Therefore, simply comparing the 6 GHz luminosity would disregard a large fraction of jet power during the giant outburst. As the higher frequencies are emitted closer to the compact object, such a scenario could imply that the electrons in the jet are first accelerated closer to the NS. 

\subsection{Future observations}
\label{sec:and_what_now}

In this paper, we have reported the first detections of radio emission from a jet during the X-ray re-brightenings of a BeXRB. Our results leave many questions unanswered, concerning both accretion in BeXRBs and the formation of jets by strongly magnetized neutron stars. To address the former, denser radio and X-ray monitoring of other giant outbursts and their decays is necessary. Such monitoring would show how common the behaviour of Sw J0243 is and whether our reasoning about the magnetospheric and corotation radius holds up under more scrutiny in other sources. Secondly, radio observations of a Type-I periastron outburst, without a prior giant outburst, can test the importance of a residual accretion flow for the formation of a jet in BeXRBs. 

To better understand jet formation from strongly magnetized neutron stars, a more extensive phenomenological picture should be constructed based on both detailed monitoring of individual sources and deep observations of a large sample of such sources. Furthermore, better spectral constraints over a large frequency range would be enlightening, especially to better probe the jet power and the (evolution of) the break frequency of these jets. Currently, such an observational campaign would require coordinated VLA/ATCA and ALMA observations \citep[similar to][for a NS LMXB]{diaztrigo18}, while in the future the proposed next-generation VLA \citep{murphy18} would be perfectly suited for such studies. 

\section*{Acknowledgements}
The authors thank the anonymous referee for reviewing this work and the VLA for rapidly accepting and performing the Director's Discretionary Time radio observations. JvdE is grateful to Eva Laplace, Lee Townsend, Alexander Mushtukov, and Sergey Tsygankov for useful discussions. JvdE, ND, and JVHS are supported by a Vidi grant from the Netherlands Organization for Scientific Research (NWO) awarded to ND. TDR is supported by a Veni grant from the NWO. ARE and RW are supported by an NWO Top grant, module 1, awarded to RW. JCAM-J is the recipient of an Australian Research Council Future Fellowship (FT140101082). GRS acknowledges support from an NSERC Discovery grant. This work made use of data supplied by the UK Swift Science Data Centre at the University of Leicester. The National Radio Astronomy Observatory is a facility of the National Science Foundation operated under cooperative agreement by Associated Universities, Inc. This research has made use of MAXI data provided by RIKEN, JAXA and the MAXI team.




\input{output.bbl}




\bsp	
\label{lastpage}
\end{document}